# A Survey of Protocols and Standards for Internet of Things

Tara Salman, Raj Jain
Department of Computer Science and Engineering
Washington University in St. Louis
{tara.salman, jain}@wustl.edu

**Abstract**

The rapid growth in technology and internet connected devices has enabled Internet of Things (IoT) to be one of the important fields in computing. Standards, technologies and platforms targeting IoT ecosystem are being developed at a very fast pace. IoT enables things to communicate and coordinate decisions for many different types of applications including healthcare, home automation, disaster recovery, and industry automation. It is expected to expand to even more applications in the future. This paper surveys several standards by IEEE, IETF and ITU that enable technologies enabling the rapid growth of IoT. These standards include communications, routing, network and session layer protocols that are being developed to meet IoT requirements. The discussion also includes management and security protocols in addition to the current challenges in IoT which gives insights into the current research to solve such challenges.

## I. Introduction

Internet of Things (IoT) is getting a high interest and popularity in both industry and academia. Its vision is to have plug-and-play smart devices that can be deployed in any environment and that can communicate and collaborate with other devices. This has become feasible only with the recent evolution in the internet protocols, sensing devices, efficient computing, big data analytics, and machine to machine (M2M) communication. According to the Gartner report in 2015, IoT technologies that enable this vision are receiving billions of dollars in investment and a high research interest, while much more is expected to come in the near future [1].

IoT is composed of two terms: "internet" and "things". It allows things, or non-computer devices, to hear, see, think, compute, and act by allowing them to communicate and coordinate with each other in decision making. In other words, it allows things to act smartly and make consensus decisions that benefits many applications. They transform objects or sensors from being passive observers to actively computing, communicating, collaborating and making critical decisions. The fundamental technologies of embedded powerful sensors, new computing paradigms, data analytics, lightweight communication, and internet protocols lead IoT to offer such important services, however, they introduce the need for specialized standards and communication protocols to handle the resulting challenges.

IoT plays a significant role in different types of applications including healthcare, transportation, automation, agriculture, vehicles and emergency response to disasters. In addition, it is expected to play additional roles to improve the quality of life, business applications, and smart homes. An example of currently available IoT ecosystem is smart homes, which are composed of sensors for controlling temperature, heat, and air conditioning in our homes remotely. Future extensions of such system can be preparing our coffee, controlling TV, tracking our health statistics and driving our vehicles. These applications would impose further challenges and need for standards to handle the diversity of application requirements.

In this paper, we present an overview of current IoT standards and protocols that are being developed for different layers of the networking stack, including: Medium Access Control (MAC) layer, network layer, and session layer. In addition to that, we highlight some of the management and security standards that are being developed for all these layers. We present standards developed by Internet Engineering Task Force (IETF), Institute of Electrical and Electronics Engineers (IEEE), International Telecommunication Union (ITU) and other standards organizations. In addition, we briefly discuss IoT current challenges and further research opportunities.

The rest of the paper is organized as follows: Section II describes the first layer of networking protocols, which is the data link layer and associated MAC protocols. Section III handles the network layer routing protocols while Section IV presents network layer encapsulation protocols. Section V handles the session layer protocols. Section VI summarizes the management protocols and Section VII describes the security mechanisms in key protocols and different security standards specialized for IoT. Section VIII highlights some discussion points about IoT challenges. Finally, Section IX summarizes our discussion and highlights the main points presented.

### A. Related Works

With the current anomalous research interest in IoT, many new protocols are being standardized every year. Hence, survey papers are continuously being written to highlight different aspects of standardization related to IoT. Examples of such papers include a survey of IETF standards in [2], security protocols in [3], and application, or transport, layer standards in [4]. Other papers discuss a specific layer of standardizations such as communication protocols or routing. Most importantly, [5] summarizes the most important standards that are offered by different standards organizations up to 2015. It also provides a discussion of different IoT challenges such as mobility and scalability. In this paper, we aim to provide a comprehensive survey of newly rising standards, drafts, and protocols that extend the work done in [5]. This allows us to discuss more standards, add some of the recent standard drafts offered in the IETF, and discuss the state of the art protocols that are expected to go for standardization in the near future.

### B. IoT Ecosystem

IoT ecosystem, as shown in Fig. 1, consists of a seven-layer model: market, acquisition, interconnection, integration, analysis, applications and services. At the bottom layer, the market layer or application domain, may be smart grid, connected home, or smart health, etc. The second layer consists of sensors and smart devices that can be considered as the application core. Sensors type and distribution varies depending on the desired applications. Examples of such sensors are temperature sensors, humidity sensors, electric utility meters, or cameras. The third layer consists of the interconnection layer that facilitates the communication of sensor data to a data center or a cloud. There the data is combined with other known data sets such as geographical data, population data, or economic data. Furthermore, the combined data are scrutinized using machine learning and data mining techniques. New application level collaboration and communication software are needed to enable such large distributed applications. Such paradigms include software defined networking (SDN), services oriented architecture (SOA), etc. Finally, the top layer consists of services that result, such as energy management, health management, education, transportation, etc. Security and management are required for each of these 7-layers that are built on top of each other, hence, they are shown on the side.

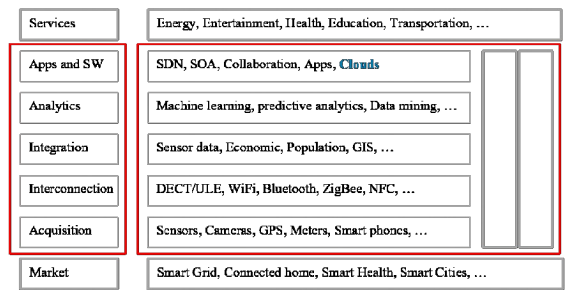

Fig. 1: IoT ecosystem

This paper focuses on the interconnection layer. This layer itself integrates multiple layers as shown in Fig. 2. These include the data link, network, and transport/session layers. The data link layer connects two IoT elements which could be two sensors or a sensor and gateway device that connects a set of sensors to the Internet. Often there is a need for multiple sensors to communicate and aggregate information before getting to the Internet. Specialized protocols have been designed for routing among sensors and are part of the network layer. The session layer protocols enable messaging among various elements of the IoT communication subsystem. In addition, several security and management protocols have also been developed for IoT as shown in the figure.

Fig. 2: Protocols for IoT

Standards to cover all those five layers were proposed by several standardization organizations. Prominent among them are IEEE, IETF, and ITU. Generally speaking, IEEE mostly works on data link, IETF work on networks and several organizations work on the session, security and managements. These protocols and many others are listed in Fig. 2. Although Fig. 2 was made as current as possible, new standards are continuously admitted and hence may appear in the future. This paper aims briefly discuss each of the ones presented in Fig. 2, however, we empathize more on protocols shown in bold face. We consider these as most commonly recommended and/or designed especially for IoT.

## II. IoT Data Link Protocols

In this section, we discuss the data link layer protocol standards. The discussion includes physical (PHY) and MAC layer protocols which are combined by most standards.

### A. IEEE 802.15.4e

IEEE 802.15.4 is a data link standard that is commonly used in the MAC layer. The standard specifies the frame format, headers, destination and source addresses and identifies how the communication between the nodes can happen. The traditional frame formats used in networking are not suitable for power constrained IoT devices. In 2008, IEEE 802.15.4e was created to extend IEEE 802.15.4 and support low power communication. It uses time synchronization and channel hopping to enable high reliability, low cost communication in IoT data links. Its specific MAC features can be summarized as follows [6]:

- **Slotframe Structure:** Scheduling and assigning nodes' state at a specific time is defined by IEEE 802.15.4e frame structure. A node can be in sleep, transmit, or receive state. When transmitting, it sends its data and waits for an acknowledgment. When receiving, the node turns on its radio before the scheduled receiving time, receives the data, sends an acknowledgement, turn off its radio, delivers the data to the upper layers and goes back to sleep. In the sleep mode, the node turns off its radio to save power and stores all messages that it needs to send at the next transmission opportunity.

- **Scheduling:** The scheduling algorithm can be defined by the designer based on application needs, however, scheduling should meet mobility and handover requirements to be accepted by the standards. Scheduling can be centralized by a manager node which is responsible for building the schedule, informing others about the schedule and other nodes will just follow the schedule.

- **Synchronization:** Nodes' synchronization is needed to maintain node connectivity to its neighbors and to the gateway. It can be done through acknowledgment-based or frame-based synchronization. In acknowledgement-based mode, the nodes that were already in communication send acknowledgments for reliability guarantees which can be used to maintain connectivity as well. In frame-based mode, the nodes are not communicating and hence, they send an empty frame at pre-specified intervals, about 30 second typically.

- **Channel Hopping:** Channel hopping was introduced in IEEE802.15.4e to allow time slotted access to the wireless medium using time slotted channel hopping (TSCH). It requires changing frequency using a pre-determined random sequence that is arbitrary in length, can go up to 511 elements, and cover all or a subset of channels that are available to the physical layer. Subsequent packets are send on different channels following the specified sequence and thus in a pseudo random hopping pattern. This introduces frequency diversity and reduces the effects of interference and multi-path fading. Furthermore, it increases security as such hoping can be a defense against selective jamming attacks.

- **Network Formation:** Advertising the network and requests to join are two important requirements for any MAC protocol. In 802.15.4e, nodes listen to advertisement commands and upon receiving at least one such command, it can send a join request to the advertising device. In a centralized system, the join request is routed to the manger node and processed there while in distributed systems, they are processed locally. Once a device joins the network and it is fully functional, the formation is disabled and will be activated again if it receives another join request.

### B. IEEE 802.11ah

IEEE 802.11ah is the least overhead version of IEEE 802.11 standards which is lightweight to meet IoT needs. IEEE 802.11 standards (also known as Wi-Fi) are the most commonly used wireless standards in traditional networking. They have been widely adopted for all digital devices, including laptops, mobiles, tablets, and digital TVs. However, the original WiFi standards are not suitable for IoT applications due to their frame overhead and high power consumption. Hence, IEEE 802.11 working group initiated 802.11ah task group to develop a standard that supports low overhead, power friendly communication suitable for sensors and motes [7]. IEEE 802.11ah MAC layer features include:

- **Synchronization Frame:** Only valid stations with valid channel information can transmit by reserving the channel medium. A station knows that it can transmit if it receives the duration field packet correctly. If it does not receive the frame correctly, then it should wait for a duration called *Probe Delay*. *Probe Delay* can be configured by the access points in 802.11ah and announced by transmitting a synchronization frame at the beginning of the transmission cycle.

- **Efficient Bidirectional Packet Exchange:** Allowing both uplink and downlink communication between access points and the sensors is a feature in IEEE 802.11ah. This feature reduces power consumption as the sensors will go to sleep as soon as they finish their communication.

- **Short MAC Frame:** IEEE 802.11ah reduces frame size from 30 bytes in traditional IEEE 802.11 to 12 bytes. Hence 802.11ah frame is much less overhead frame and more suitable for IoT application.

- **Null Data Packet:** Traditional 802.11 standards had acknowledgment (ACK) frames of 14 bytes with no data. Such feature would add lots of overhead, especially for IoT. 802.11ah solves this problem by introducing a tiny signal, called preamble, which is used in place of ACKs and is much less in size.

- **Increased Sleep Time:** As this standard is designed for power constrained devices, it allows a long sleep period and waking up infrequently to exchange data only.

### C. WirelessHART

WirelessHART is a MAC layer standard that works on top of IEEE 802.15.4 PHY and uses time division multiple access (TDMA) in its MAC. It uses advanced encryption algorithms to encrypt messages and check for integrity. Hence, it is more secure and reliable than others. Its architecture, as shown in Fig. 3, consists of a network manager, a security manager, a gateway to connect the wireless network to the wired networks, wireless devices as field devices, access points, routers and adapters. The standard offers end-to-end, per-hop or peer-to-peer security mechanisms. End to end security mechanisms enforce security from sources to destinations while per-hop mechanisms secure it to next hop only [8], [9].

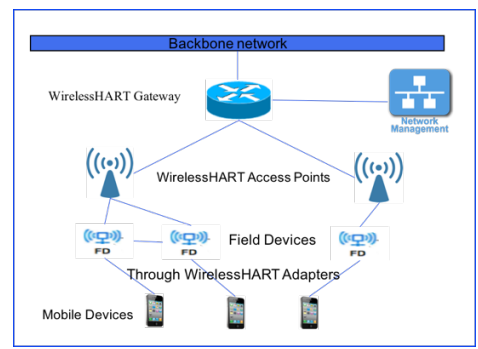

Fig. 3: WirelessHART Architecture

### D. Z-Wave

Z-Wave is a low power consumption MAC standard that was designed for home automation, but recently used in many IoT applications, including smart homes and small commercial domains. It covers up to 30-meter distance, point-to-point communication and is suitable for small messages. It uses CSMA/CA for media access in addition to small ACK messages for reliable transmission. It follows a master/slave architecture in which the master controls the slaves, sends them commands, and handles scheduling of the whole network [10].

### E. Bluetooth Low Energy

Another short-range communication standard for data link layer that is widely used in IoT is Bluetooth low energy, or Bluetooth smart. It is mostly used in in-vehicle networking. It has a small latency that is 15 times smaller than original Bluetooth standards. Its low energy can reach ten times less than the classic Bluetooth. Its access control uses a contention-less MAC with low latency and fast transmission. It adopts a master/slave architecture and offers two types of frames: adverting and data frames. The advertising frame is used for discovery and is sent by slaves on one or more of dedicated advertisement channels. Master nodes sense advertisement channels to find slaves and connect them. After connection, the master tells the slave it's waking cycle and scheduling sequence. Nodes are awake usually only when they are communicating and they go to sleep otherwise to save their power [11], [12].

### F. ZigBee Smart Energy

ZigBee is one of the most commonly used standards in IoT that is dedicated for medium-range communication in smart homes, remote controls, and healthcare systems. Its networking topologies include star, peer-to-peer, or cluster-tree. A coordinator controls the network and is located at the center in a star topology, the root of a tree or cluster topology and anywhere in the peer-to-peer topology. The ZigBee standard defines two stack profiles: ZigBee and ZigBee Pro. These stack profiles support full mesh networking and work with different applications allowing implementations with low memory and processing power. ZigBee Pro offers more features including security using symmetric-key exchange, scalability using stochastic address assignment, and better performance using efficient many-to-one routing mechanisms [13].

### G. DASH7

DASH7 is a new wireless communication protocol that is used for active RFID devices, operates in globally available industrial scientific medical (ISM) band. It is mainly designed for scalable, long-range outdoor coverage with a higher data rate compared to traditional ZigBee. It is a low-cost solution that supports encryption and IPv6 addressing. It supports a master/slave architecture and is designed for burst, lightweight, asynchronous and transitive traffic and, thus, suitable for IoT. Its MAC layer features can be summarized as follows [14]:

- **Filtering:** An incoming frame is filtered by three processes: cyclic redundancy check (CRC) validation, a 4-bit subnet mask, and a link quality assessment. If the frame passes those checks, it can be processed, but otherwise, it will not.
- **Addressing:** Two types of addresses are used; the unique identifier which is the EUI-64 ID and dynamic network identifier which is a 16-bit address specified by the network administrator.
- **Frame format:** A variable length MAC frame that can be 255 bytes at maximum including addressing, subnets, estimated power of the transmission and some other optional fields.

### H. HomePlug

HomePlug GreenPHY (HomePlugGP) is a MAC protocol developed by the HomePlug Powerline Alliance and is mostly used in home automation applications. HomePlug package, including HomePlug-AV, HomePlug-AV2, covers both PHY and MAC layers of the networking stack. HomePlug-AV is the basic power line communication protocol, which uses TDMA and CSMA/CA as MAC layer protocols, supports adaptive bit loading which allows it to change its rate depending on the noise level and uses orthogonal frequency division multiplexing (OFDM) and four modulation techniques.

HomePlugGP is designed for IoT applications such as smart home and smart grid applications. It is basically designed to reduce the cost and power consumption of HomePlug-AV while keeping its interoperability, reliability and coverage. Hence, it uses OFDM, as in HomePlug, but with one modulation only. In addition, it utilizes robust OFDM coding to support low rate and high transmission reliability. HomePlug-AV uses only CSMA as a MAC layer technique while HomePlugGP uses both CSMA and TDMA. Moreover, HomePlugGP has a power-save mode that allows nodes to sleep by synchronizing their sleep time and waking up when necessary [15].

### I. G.9959

This is an ITU MAC layer standard that is designed for low bandwidth, low cost, and half-duplex reliable wireless communication. It is dedicated for delay sensitive applications where time is critical, reliability is important, and low power consumption is required. The MAC layer characteristics include: unique network identifiers that allow up to 232 nodes to join one network, collision avoidance mechanisms, back-off time in case of collision, automatic retransmission to guarantee reliability, dedicated wakeup pattern that allows nodes to sleep when they are out of communication and hence save their power. G9959 MAC layer features include unique channel access, frame validation, ACK, and retransmission [16], [17].

### J. LTE-A

Long-term evolution advanced (LTE-A) is a collection of cellular networking standards that is designed to meet M2M and IoT requirements in such networks. It is one of the most scalable and cost effective standards compared to other cellular protocols. LTE-A was started in 2009 with multiple releases that are continuously coming to support new technologies. It traditionally uses orthogonal frequency division multiple access (OFDMA) as a medium access technology, in which the frequency is divided into multiple subcarriers. The architecture of LTE-A consists of a core network (CN), a radio access network (RAN), and the mobile nodes. The CN is responsible for controlling mobile devices and to keep track of their IPs. RAN is responsible for establishing the control and data planes and handling the wireless connectivity and radio-access control. RAN and CN communicate using S1 link, as shown in Fig. 4 where RAN consists of the eNB's to which other mobile nodes are connected wirelessly [18].

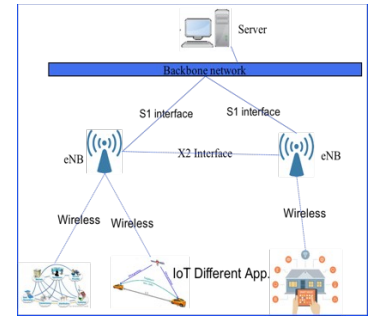

Fig. 4: LTE-A Architecture

Moreover, the new releases of LTE-A (LTE Rel-13 and Rel-14) introduce new features that were designed to fit the upcoming 5G requirements [72]. Rel-13 introduced three major features: FD-MIMO, enhanced spectrum and carrier aggregation, and new services for machine type commination. Full dimension multiple input multiple output (FD-MIMO) aims to increase spectrum efficiency using large number of antenna ports at the base station. The utilization of additional frequency resources is done by using unlicensed spectrum in addition to the already used licensed spectrum frequencies. In this way, more frequencies are used and it keeps backward compatibility with existing devices. Further, increasing the peak rate and efficient frequency resources distribution were used to enhance carrier aggregation in LTE-A. Moreover, LTE-A Rel-13 offered new services for machine type communication including cost reduction, support of extended coverage, indoor positioning, and broadcast and multicast support in a single cell.

LTE Rel-14 specification is expected to further enhance FD-MIMO with more antenna porta, robust transmission and reduced feedback. In addition, the release is expected to standardize latency reduction, vehicle to anything and downlink multi-user transmission which were discussed as feasibility studies in Rel-14 [73].

### K. LoRaWAN

LoRaWAN is a newly developed long-range wide-area network wireless technology designed for IoT applications with power saving, low cost, mobility, security, and bidirectional communication requirements. It is a low-power consumption optimized protocol designed for scalable wireless networks with millions of devices. It supports redundant operation, location free, low cost, low power and energy harvesting technologies to support the future needs of IoT while enabling mobility and ease of use features [19].

### L. Weightless

Weightless is another newly developed wireless technology for the IoT MAC layer that is provided by the Weightless special interest group (SIG) - a non-profit global organization. Two standards can be used: Weightless-N and Weightless-W. Weightless-N was the first standard developed to support IoT requirements using TDMA with frequency hopping to minimize the interference. It uses ultra-narrow bands in the sub-1GHz ISM frequency band. On the other hand, Weightless-W provides the same features, but uses television band frequencies [20].

### M. DECT/ULE

DECT (digital enhanced cordless telecommunications) is a universal European standard that is designed for cordless phones. Recently, they provided an extension, called DECT/ULE (ultra-low energy), which specifies a low-power and low-cost air interface technology that can be used for IoT applications. It has a dedicated channel assignment and, hence, has much more tolerance to interference and congestion problem. DECT/ULE supports FDMA, TDMA and time division multiplexing, which were not supported in the original DECT protocol [21].

### N. EnOcean

EnOcean is an energy saving wireless technology that is primarily used for automation, but can be used for other IoT applications. The basic idea is to use efficient harvesting of motion, or any type of environmental energy, and convert it to usable energy using converters. This protocol has a relatively low packet size and is mostly used in heating, ventilation, and air conditioning IoT applications [22].

### O. Others

In addition to all the previously discussed data link protocols, near field communication (NFC), ANT, and International Society of Automation (ISA100.11a) can be used as well. These standards are rarely used in IoT as they are getting incomparable to the new upcoming ones, which we discussed in this section. NFC is used for short range communication in an ad-hoc manner. It operates at relatively low frequencies and uses radio frequency identifier to power on the receiver and start the peer-to-peer communication [23]. ANT is a multi-cast wireless protocol that operates in a master-slave manner. It is mostly used for wireless sensor networks, operates at the 2.4 GHz frequencies and conceptually similar to Bluetooth low energy [24]. ISA100.11a is the ISA standards developed for wireless networking in industrial automation control [25].

### P. Summary

In this section, various data link protocols were briefly discussed with their main differences and usefulness in IoT medium access. Those protocols are mostly standardized by IEEE, ITU or other wireless standards organizations. Generally speaking, the most widely used standards in IoT are Bluetooth and ZigBee. IEEE 802.11ah, on the other hand, is the most compatible one with IEEE 802.11 which is the most used infrastructure in other wireless applications. Furthermore, some providers and IoT markets would seek for more reliable and secured technology and hence would use HomePlug for LAN connectivity. Newly arising LoRaWAN seems to be promising for outdoor applications.

## III. Network Layer Routing Protocols

In this section, some standards and protocols for IoT routing are briefly discussed. It should be noted that this paper divides the network layer in the networking stack into two sublayers: routing layer which handles the transfer of packets from source to destination, and an encapsulation layer that forms the packets. Encapsulation standards will be discussed in the next section.

### A. RPL

Routing protocol for low-power and lossy networks (RPL) is a distance-vector protocol designed at IETF for routing in IoT system. It supports all the previously discussed MAC layer protocols and some other protocols that are not designed for IoT. It is based on destination-oriented directed-acyclic graphs (DODAG) that have only one route from each leaf node to the root through which all the traffic from the leaf node will be routed to. Initially, each node sends a DODAG information object (DIO) advertising itself as the root. DIO is propagated on the network and the whole DODAG is gradually built. When communicating, a destination advertisement object (DAO) is sent from the node to its parents, propagated to the root, and the root decides where to route it depending on the destination. New nodes who wish to join the network sends a DODAG information solicitation (DIS) request for joining and the root will reply with a DAO acknowledgment (DAO-ACK) confirmation. RPL nodes can be stateless, which is most common, or stateful. A stateless node keeps track of its parents only. Only root has the complete knowledge of the entire DODAG. Hence, all communications go through the root. A stateful node keeps track of its children and parents and hence when communicating inside a sub-tree of the DODAG, it does not have to go through the root [26].

### B. CORPL

Cognitive RPL, CORPL, is a protocol that extends RPL and uses the same DODAG technology, but with a couple of modifications to RPL. First, it introduces opportunistic forwarding which enables the packet to have multiple forwarders set but only the best next hop will be chosen to forward the packet. Then, each node will maintain a forwarding list instead of its parent only and updates its neighbor with its changes using DIO messages. Based on the updated information, each node dynamically updates its neighbor priorities in order to construct the forwarders set [27].

### C. CARP and E-CARP

Channel-aware routing protocol (CARP) is another routing protocol that is based on distributed networks and designed for underwater communication. It is a lightweight packet forwarding protocol and, hence, can be applied to IoT systems. It considers historical link quality measurements to select the forwarding route. Network initialization and data forwarding are the two scenarios that should be considered in such protocols. In network initialization, a HELLO packet is broadcasted from the sink to all other nodes in the networks. In data forwarding, the packet is routed from sensor to sink in a hop-by-hop fashion. Each next hop is determined independently.

The main problem with CARP is that it does not support the reusability of previously collected data. In other words, if the application requires sensor data only when it changes significantly, then CARP data forwarding is not beneficial to that specific application. An enhancement of CARP was done in E-CARP by allowing the sink node to save previously received sensory data. When new data is needed, E-CARP sends a ping packet which is replied with new data from the sensor nodes. Thus, E-CARP reduces the communication overhead drastically [28].

### D. Summary

This section discussed three routing protocols that can be used in IoT routing sublayers. RPL is the standardized distance vector protocol and, most commonly used one. CORPL is a non-standard extension of RPL that is designed for cognitive networks and utilizes the opportunistic forwarding to forward packets at each hop. On the other hand, E-CARP is the only distributed link quality measurement based routing protocol that is designed for IoT sensor network applications. E-CARP is used for underwater communications mostly. Since it is not standardized, it is not yet used for other IoT applications.

## IV. Network Layer Encapsulation Protocols

Addressing IoT devices with IPv6 long addresses and how they can fit in small, lightweight IoT data link frames were challenges that needed to be taken care of through standards. Hence, IETF is developing a set of frame formatting standards to encapsulate IPv6 datagrams in different small data link frames to be used in IoT applications. In this section, we review these standards briefly.

### A. 6LoWPAN

IPv6 over low power wireless personal area network (6LoWPAN) is one of the first and extensively used IETF standards in this category. It efficiently encapsulates IPv6 long headers in IEEE802.15.4 small MAC frames, which cannot exceed 128-byte length. 6LoWPAN specifications allow many features including: different length addresses, different networking topologies, low bandwidth, low power consumption, cost efficient, scalable networks, mobility, reliability, and long sleep times. Header compression is used in the standards to reduce transmission overhead, fragmentation to meet the 128-byte maximum frame length in IEEE802.15.4, and support of multi-hop delivery. Frames in 6LoWPAN use four types of headers: No 6loWPAN header (00), dispatch header (01), mesh header (10) and fragmentation header (11). In No 6loWPAN header case, any frame that does not follow 6loWPAN specifications is discarded. Dispatch header is used for multicasting and IPv6 header compressions. Mesh headers are used for broadcasting; while fragmentation headers are used to break long IPv6 header to fit into 128-byte fragments.

### B. 6TiSCH

6TiSCH is another IETF standard designed by 6TiSCH working group. It specifies ways to pass long IPv6 headers through TSCH mode of IEEE 802.15.4e data links. This mode stores the available frequencies and their time slots in a matrix called channel distribution usage matrix. This matrix is divided into multiple chunks where each chunk contains time and frequencies and is globally known to all nodes in the network. Within the same interference domain, nodes coordinate and negotiate their scheduling such that they all get to transmit without interruptions. Scheduling becomes an optimization problem where time

slots are assigned to a group of neighboring nodes sharing the same application. The standard does not specify how the scheduling can be done and leaves that to be an application specific problem in order to allow for maximum flexibility for different IoT applications. The scheduling can be centralized or distributed depending on the application or the topology used in the MAC layer [29].

### C. 6Lo

A newly assigned IETF group called IPv6 over networks of resource-constrained nodes (6Lo) is working to propose a set of standards on transmission of IPv6 frames on various data links. Even though 6LowPAN and 6TiSCH were developed for encapsulation purposes, it became clear that more standards are needed to cover all data link standards. Therefore, 6Lo was formed by IEFT for this purpose. At the time of this writing most of the 6Lo specifications have not been finalized and are in various stages of drafts. For example, IPv6 over IEEE 485 Master-Slave/Token Passing (MS/TP) networks, IPV6 over DECT/ULE, IPV6 over NFC, IPv6 over IEEE 802.11ah, and IPv6 over wireless networks for industrial automation process automation (WIA-PA) drafts are being developed to specify how to transmit IPv6 datagrams over their respective data links [30]. Two of these 6Lo specifications" IPv6 over G.9959" and" IPv6 over Bluetooth Low Energy" have been approved as an RFC and are described next.

### D. IPv6 over G.9959

This standard, defined in IETF RFC 7428, specifies the frame format for transmitting IPv6 packets on G.9959 data links, discussed in Subsection II-I above. In G.9959, a unique 32-bit home network identifier is assigned by the controller and 8-bit host identifier that is allocated for each node. Hence, an IPv6 link local address must be constructed by the link layer derived 8-bit host identifier so that it can be compressed in G.9959 frame. Furthermore, the same header compression as in 6lowPAN is used here to fit an IPv6 packet into G.9959 frames. It should be noted that RFC 7428 has a security feature by allowing a shared network key that is used for encryption. However, this is not enough for security critical applications which need to have end-to-end encryption and authentication and that is mostly handled by other protocols and higher layer security mechanisms [30].

### E. IPv6 over Bluetooth Low Energy

RFC 7668 [31] specifies the format of IPv6 over Bluetooth low energy, which was discussed in Subsection II-E. It reuses most of the 6LowPAN compression techniques. Fragmentation is done at the logical link control and adaptation protocol (L2CAP) sublayer in Bluetooth. Thus, the fragmentation feature of 6LowPAN is not used here. Further, Bluetooth low energy does not currently support formation of multi-hop networks at the link layer. Instead, a central node acts as a router between lower-powered peripheral nodes. Thus, multi-hop feature in 6LowPAN is not used as well.

### F. Summary

This section discussed encapsulating long IPv6 datagrams into small MAC frames for IoT. First, 6LowPAN and 6TiSCH for IPv6 over 802.15.4 and 802.15.4e were discussed. Such protocols are important as 802.15.4e is the most widely used encapsulation framework designed for IoT. Following that, 6Lo specifications are briefly and broadly discussed just to present their existence in IETF standards. These drafts handle transmitting IPv6 datagrams over different channel access mechanisms using 6LoWPAN standards. Then, two of 6Lo specifications, which have become IETF RFCs, are discussed in more detail. The importance of presenting these standards is to highlight the challenge of interoperability between different networking stack layers which is still challenging due to the diversity of data link protocols.

## V. Session Layer Protocols

In this section, we review several IoT session layer protocols that are used for message passing and that have been standardized by different standardization organizations. At the transport layer, TCP and UDP are the dominant protocols for most of the applications, including IoT. However, several message distribution functions are required depending on IoT application requirements. It is desirable that these functions be implemented in interoperable standard ways. These are the so called "Session Layer" protocols which are described in this section.

## A. MQTT

Message queue telemetry transport (MQTT) is a 2013 standard from the Organization for the Advancement of Structured Information Standards (OASIS). It was introduced back in 1999 by IBM [32], [4]. It provides the connectivity between applications and users at one end, network and communications at the other end. It is a publish/subscribe architecture, as shown in Fig. 5, where the system consists of three main components: publishers, subscribers, and a broker. In IoT, publishers are the lightweight sensors that connect to the broker to send their data and go back to sleep whenever possible. Subscribers are applications that are interested in a certain topic, or sensory data, so they connect to brokers to be informed whenever new data are received. The brokers classify sensory data in topics and send them to subscribers interested in those topics only.

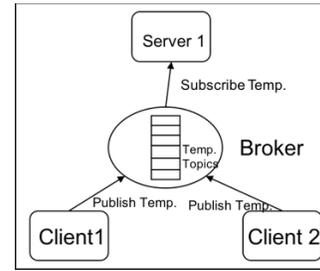

Fig. 5: MQTT Architecture

## B. SMQTT

An extension of MQTT, secure MQTT (SMQTT), was proposed in [33] to provide a lightweight attribute based encryption. Such encryption uses a multicast feature, in which one message is encrypted and delivered to multiple other nodes, which is quite common in IoT applications. Generally, the algorithm consists of four main stages: setup, encryption, publish and decryption. In the setup phase, the subscribers and publishers register themselves to the broker and get a master secret key according to their developer's choice of key generation algorithm. Then, when the data is published, it is encrypted and published by the broker which sends it to the subscribers. Finally, it is decrypted at the subscribers which have the same master secret key. The key generation and encryption algorithms are not standardized.

## C. AMQP

Advanced message queuing protocol (AMQP) is another OASIS standard that was designed for the financial industry, runs over TCP, and uses publish/subscribe architecture, similar to MQTT. The main difference in these standards is that the broker is divided into two main components: exchange and queues, as shown in Fig. 6. The exchange component is responsible for receiving publisher messages and distributing them to queues following pre-determined roles. Subscribers connect to those queues, which basically represent the topics, and get the sensory data whenever they are available [34].

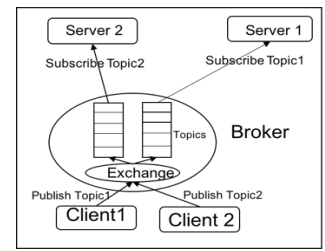

Fig. 6: AMQP Architecture

## D. CoAP

Another session layer protocol designed in the IETF constrained RESTful environment (Core) working group is constrained application protocol (CoAP), which is designed to provide low overhead RESTful (HTTP) interface. Representational state transfer (REST) is the standard interface which is extensively used in today's web applications. However, REST has a significant overhead and power consumption which made it unsuitable for IoT platforms. CoAP is designed to solve the REST problems and enable IoT applications to use RESTful services while meeting their requirements. It is built over UDP, instead of TCP, and has a lightweight mechanism to provide reliability. CoAP architecture is divided into two main sublayers: messaging and request/response. The messaging sublayer is responsible for the reliability and duplication of messages while the request/ response sublayer is responsible for communication.

As shown in Fig. 7, CoAP can have four messaging types: confirmable, non-confirmable, piggyback, and separate. Confirmable and non-confirmable represent the reliable and unreliable transmissions, respectively while the other modes are used for request/response. Piggyback is used for client/server direct communication, where the server sends its response directly after receiving the message, i.e., within the acknowledgment message.

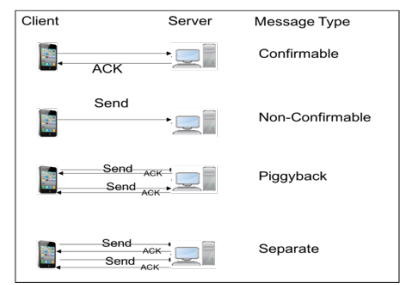

Fig. 7: CoAP Messages

On the other hand, the separate mode is used when the server response comes with a message separate from the acknowledgment, and may take some time to be sent by the server. As in HTTP, CoAP utilizes get, put, push, delete message requests to retrieve, create, update, and delete, respectively [35], [4].

### E. XMPP

Extensible messaging and presence protocol (XMPP) is a protocol that was originally designed for chats and messages exchange applications. It is based on XML language and was standardized by IETF more than a decade ago. It is quite popular and is highly efficient when used over the internet. Recently, its usage was extended for IoT and SDN applications due to the standardized use of XML which makes it easily extensible. XMPP supports both publish/subscribe and request/ response architecture and it is up to the application developer to choose which architecture to use. It is designed for near real-time applications and, thus, efficiently supports low-latency small messages. It does not provide any quality of service guarantees and, hence, is not practical for M2M communications. Moreover, XML messages create additional overhead due to lots of headers and tag formats which increase the power consumption that is critical for IoT application. Hence, XMPP is rarely used in IoT but there is some interest in enhancing its architecture to support IoT applications [36], [4].

### F. DDS

Data distribution service (DDS) is a messaging standards designed by the Object Management Group (OMG). It uses a publish/subscribe architecture and mostly used for M2M communications [37]. The highest beneficial features of this protocol are the outstanding quality of service levels and the reliability with the use of a broker-less architecture, which suits IoT and M2M communication. It offers 23 quality-of-service levels which allow it to offer a variety of quality criteria, including: security, urgency, priority, durability, reliability, etc. It defines two sublayers: data-centric publish-subscribe and data-local reconstruction sublayers. The first takes the responsibility of message delivery to the subscribers while the second is optional and allows a simple integration of DDS in the application layer. Publisher layer is responsible for sensory data distribution. Data writer interacts with the publishers to agree about the data and changes to be sent to the subscribers. Subscribers are the receivers of sensory data to be delivered to the IoT application. Data readers basically read the published data and deliver it to the subscribers and the topics are basically the types of data that are being published. In other words, data writers and data reader take the responsibilities of the broker in the broker-based architectures.

### G. Summary

Several IoT session layer standards and protocols have been proposed in the literature and were briefly discussed in this section. These standards are totally application dependent and the choice among them depends on the desired application. MQTT is the most widely used in IoT due to its low overhead and power consumption. The choice among these standards is organizational and application specific. For example, if an application has already been built with XML and can, therefore, accept a bit of overhead in its headers, XMPP might be the best option to choose among session layer protocols. On the other hand, if the application is overhead and power sensitive, then choosing MQTT would be the best option. However, that comes with the additional broker implementation. If the application requires REST functionality as it will be HTTP based, then CoAP would be the best option if not the only one. Table I summarizes comparison points between these different session layer protocols.

Table I: A Comparison of IoT Session Layer Standards

| Protocols | UDP/TCP | Architecture | Security and QoS | Header Size (bytes) | Max Length (bytes) |
|---|---|---|---|---|---|
| **MQTT** | TCP | Pub/Sub | Both | 2 | 5 |
| **AMQP** | TCP | Pub/Sub | Both | 8 | - |
| **CoAP** | UDP | Req/Res | Both | 4 | 20 (typical) |
| **XMPP** | TCP | Both | Security | - | - |
| **DDS** | TCP/UDP | Pub/Sub | QoS | - | - |

## VI. IoT Management Protocols

In this section, we provide an overview of several management protocols used in IoT to provide heterogeneous device management and communication. We start by discussing two protocols to handle data link heterogeneity. Then, we discuss a few remote device management protocols that are used mostly in M2M and IoT applications. Management protocols play a significant role in IoT due to the diversity and the requirement at different layers of networking. The need for heterogeneous and easy communication between protocols at the same or different layers is critical for IoT applications. Existing standards mainly facilitate communication between protocols at the same layer and it is still a challenge to facilitate communication at different layers in IoT.

### A. IEEE 1905.1 - Interconnection of Heterogeneous Data links

As was shown in Section II, IoT has many different and diverse MAC layer protocols and, hence, interoperability among these standards is critical. This standard, designed in IEEE, would handle such interoperability by providing an abstraction layer that is built on top of all these heterogeneous MAC protocols [38]. This abstraction allows different protocols to communicate by hiding their diversity without requiring any change to their design. The abstraction layer allows the exchange of messages, called control message data units (CMDUs), among all standards compatible devices. As shown in Fig. 8, all IEEE 1905.1 compliant devices understand a common "abstraction layer management entity (ALME)" protocol, which offers different services including: neighbor discovery, topology exchange, topology change notification, measured traffic statistics exchange, flow forwarding rules, and security associations.

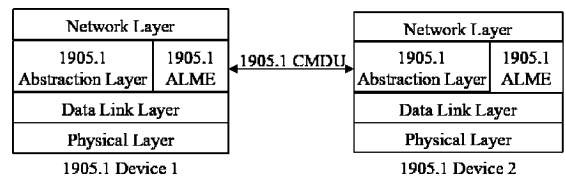

Fig. 8: IEEE 1905.1 Protocol Structure

### B. Smart Transducer Interface

The smart transducer interface is another standard that is provided by IEEE 1451 and used to facilitate the management of different analog transducers and sensors. The idea of this interface is to use plug and play identification by standardized transducer electronic data sheets (TEDSs). Each transducer contains a TEDS which includes all the information needed by the measurement system, including device ID, characteristics, and interface. Data sheets are stored in embedded memory within the transducer or the sensor and have a defined encoding mechanism to understand a broad number of sensor types and applications. The memory usage is minimized by utilizing small XML based messages which are understood by different manufactures and different applications [39].

### C. TR-069

Technical report 069, designed by the Broadband Forum and entitled "CPE (customer-premises equipment) WAN management protocol (CWMP)," is an industry specification designed for remote management of M2M devices by HTTP messages. In this specification, the management is done by HTTP messages sent from server to clients or desired devices. The specification is critical for M2M devices, but since it relies on HTTP message, it is currently not widely used in IoT applications [40].

### D. OMA-DM

OMA device management is a protocol designed by the Open Mobile Alliance (OMA). It is used for remote provisioning, updating and managing faulty issues of M2M devices. It uses XML messages for communication, build over HTTP, and it can be applied to any XML based transport protocol, such as XMPP. However, messages in such protocol are still complex for resource-constrained IoT devices to use [41].

### E. LWM2M

Lightweight M2M is another OMA protocol that is specifically designed for IoT device management. It is a client-server protocol in which JASON (JavaScript Object Notation) messages are used for communication. It is mostly built on CoAP but can be applied to other session protocols. This protocol is used to manage device functions over the network, transfer data from the server to devices and can be extended to many IoT server-client messages [42].

F.  Summary

In this section, we discussed several management protocols for interoperability and heterogeneity of different IoT protocols. IEEE-1905.1 is used for heterogeneity of IoT MAC layer protocols, while IEEE 1451 is used for transducers and sensor management. TR-069, OMA-DM, and LWM2M are used for remote management protocols where LWM2M is more suitable and widely used for IoT. Management between protocols at different layers of IoT communication is still a challenge to be solved.

VII.  Security of IoT Protocols

Providing security for IoT platforms is another challenge to consider at all layers of networking discussed in the previous sections. Conventional security mechanisms such as cryptography and public key infrastructure sound impractical with IoT platforms due to their complexity and resource consumption. Hence, new standards are being developed with lightweight security designs. In addition, there exist a few IoT standards specializing in providing security as shown earlier in Fig. 2. In this section, we discuss several of these security standards, drafts and research works. We refer the reader to [3] for more details on IoT security standards.

A.  Security within IoT Protocol Layers

IoT security threats span all layers including data link, network, session, and application layers which motivates standards, discussed in this paper, to include security within their design. Protocols such as 802.15.4e, WirelessHART, 6LoWPAN and RPL offer some security features to secure the communication in their respective layers.

MAC 802.15.4e offers different security modes by utilizing the "security enabled bit" in the frame control field in the header. Security requirements include confidentiality, authentication, integrity, access control mechanisms and secured time-synchronized communications.

The WirelessHART standard provides robust security features by utilizing the latest and widely used security techniques. Such techniques include unique security for each message by AES-128 encryption, data integrity, and authentication, channel hopping for protection, indication of failed data access, and reports on message integrity and authentication failures. Thus, it provides different levels of security, depending on the application, using latest employed techniques.

A few IETF documents that are relevant to 6LoWPAN discuss security threats and requirements of 6LoWPAN and propose solutions. For example, RFC 4944 discusses the possibility of duplicate EUI-64 interface addresses which are supposed to be unique [43]. RFC 6282 discusses security issues that are raised due to the problems introduced in RFC 4944 [44]. RFC 6568 addresses possible mechanisms to adopt security within constrained wireless sensor devices [45]. In addition, a few recent drafts [30] discuss mechanisms to achieve security in 6loWPAN. See also [46], [47].

RPL offers different security levels using the "Security" field in its header. Information in this field indicates the level of security and the cryptography algorithm used to encrypt the message. RPL offers support for data authenticity, semantic security, protection against replay attacks, confidentiality and key management. Levels of security in RPL include: unsecured, preinstalled, and authenticated. RPL threats include selective forwarding, sinkhole, Sybil, hello flooding, wormhole and Denial of Service attacks. RFC 7416 [48] discusses security threats and possible attacks to RPL, including confidentiality, availability and integrity attacks with their possible countermeasures.

B.  TLS/DTLS

Transport layer security (TLS) and datagram TLS (DTLS) are two widely-used standards for security. They mainly provide authentication, integrity and confidentiality at the transport layer, especially used in CoAP protocols. TLS provides the security services over TCP transmission while DTLS provide such services over UDP or datagram transmission. TLS and DTLS are composed of two sublayers of protocols - record and the handshaking - which are responsible for the encapsulation and authentication, respectively. RFC 7925 discusses the detailed mechanisms used in these standards to provide security and privacy [49]. These standards can provide credentials, signature, error handling using traditional security mechanisms, but modified to fit resource constrained devices used in IoT.

### C. Ubiquitous Green Community Control Network Security

IEEE 1888.3 standards specify security requirements and mechanisms for ubiquitous green community control network protocol. These networks provide high quality, energy saving and secure mechanisms that are suitable for IoT. Security requirements include information protection, integrity, confidentiality, authentication and access control. The standard provides recommended architectures and components that are needed to achieve security in such system. More importantly, the standard specifies communication sequence and security mechanisms that can be used, including: handshaking, authentication and access control mechanisms [50].

### D. TCG

Trusted computing group (TCG) provides a guideline for secure IoT diverse applications using different use cases and security mechanisms. The mechanisms include authentication using unique identifiers, protection against middleware infections using TLS, and proven availability, confidentiality and integrity using various techniques. The techniques include the root of trust for update (RTU) and trusted platform module (TPM) which are used in TCG compatible devices [51]. Specifications can help IoT developers to pick mechanisms that secure their applications, however, it is up to the developer to balance system security with complexity and resource needs.

### E. OAuth 2.0

OAuth is an authorization framework specified in IETF RFC 6749. It enables trusted third party servers to control access rights and permissions to resources. The specification enables clients to request authorization, access from owners through an authorization server. Such server would check client credentials and access rights and decide on the permission based on such information. Messages in this framework are based on HTTP which is rarely used for IoT due to its overhead compared to others [52]. RFC 6819 [53] describes additional security considerations that extend OAuth to include new threat models and solutions. The authors in [54] discuss threats and open security issues that go beyond OAuth 2.0 and need to be solved in future versions of this protocol. These threats include credential leakage, injections, and risks in the third-party authorization server.

### F. SASL

Simple authentication and security layer (SASL) is another security framework by IETF for supporting authentication in IoT applications through servers. It decouples the application from authentication process and uses simple messages to authenticate clients using application specific authentication mechanisms. Typically, in IoT, this framework is supported by session layer protocols that support TLS and SSL, such as, MQTT and AMQP [55].

### G. ACE

Authentication and authorization in constrained environments (ACE) is a security mechanism that is designed for resource constrained devices and, hence, can be used in IoT platforms. It is conceptually like OAuth. However, it is built on CoAP based messages which is more suitable for IoT. It should be noted that the specifications have recently been approved in IETF RFC 7744 [56], and another draft [57] is in progress.

### H. Recent IETF drafts on IoT security

Despite the large amount of security protocols and standards that have been proposed for IoT, threats and malicious behaviors are still challenges requiring further research. Some of these challenges and security requirements are discussed in several recent IETF drafts which we summarize next.

Different security aspects and requirements for IoT are discussed in [58]. The discussion includes the lifecycle of IoT devices and mechanisms to provide security at bootstrapping, operation, updates and end of life stages of the device. They provide different IoT profiles, or use cases, and discuss the available security protocols for such profiles at different stages of the device lifecycle. Furthermore, they discuss the challenges of the current protocols in providing end-to-end IoT security.

The authors in [59] summarize many technical and non-technical challenges in IoT security. Such challenges were gathered from enterprises offering IoT platforms and, hence, they are practical challenges rather than research challenges. Security and privacy of data are considered as the biggest challenges faced in the current IoT implementations and, hence, lightweight solutions for these challenges are yet to be developed.

Further, another draft [60] discusses the current practices for securing IoT devices from the network perspectives. The discussion includes requirements for IoT security, the use of security protocols to provide such requirements and the challenges faced in using such protocols. In addition, the authors provide recommendations for security solutions and implementation guidelines that are helpful for IoT enterprises. Hence, this document can serve as a guideline for IoT security minimum requirements and the current security issues to be solved in further research.

I. Other work done on IoT security

In line with all the standardization work discussed in this section, a lot of research papers and surveys have been published but not yet standardized. We briefly discuss some of these in this subsection.

New protocols to provide security are discussed in [61, 62, 63]. A lightweight end-to-end key mechanism for resource constraint devices used in IoT is proposed in [61]. This protocol is based on offloading the computationally complex cryptographic operations to a trusted non-resource constrained neighbor device or node that provides strong encryption and authentication features. However, requiring a trusted third party which does all the work defeats the privacy constraint. An architectural design was proposed in [62] to meet the security and privacy requirements during the lifecycle of an IoT device. This architecture is based on an architectural reference model (ARM) which is designed by IoT-A European project for broader interoperability among IoT systems. The paper describes the instantiation, implementation, deployment, and testing of this architecture in IoT platforms. In addition, a monitoring solution for 6LoWPAN-based sensors that enables data capture, event extraction, statistics collection, analysis and reporting of wireless sensors behaviors is described in [63]. Having such reporting of the sensors would result in better intrusion detection and deep inspection on the network traffic.

Surveys of standardized and non-standardized IoT protocols are presented in [3, 64, 65]. The first survey provides an extensive analysis of IoT security protocols and mechanisms in addition to open research issues and challenges faced in this field [3]. Security challenges faced by industries offering IoT platforms are discussed in [64]. The discussion includes challenges in using the protocols discussed in the paper. Specifically, the authors focused on different security mechanisms for key generation and analyzed the effect of using those mechanisms on MQTT. Finally, several state of the art mechanisms are discussed in [65] to provide cryptographic solutions, vulnerabilities detection and intrusion identification in IoT platforms.

J. Blockchains for IoT security

A newly arising area of research in IoT security is the use of Blockchains in building smart contracts and security protection of IoT platforms. Blockchain is the distributed ledger technology that provides security by design without referring to a centralized or trusted third party authority [66]. It is traditionally used in Bitcoins, and other virtual cryptocurrency platforms, but recently been investigated in many other domains including IoT. IBM and other IoT enterprises are considering providing Blockchain solutions for IoT security [67]. Blockchains can be also used to provide privacy provisioning in IoT platforms [68]. In [69], the authors discussed mechanisms to share data between IoT devices and organizations in a secured way using blockchains. In addition, building smart contracts using blockchain technology has been discussed in [70]. Furthermore, [71] provided a survey of blockchain based architectures built for IoT platforms.

K. Summary

Security is still one of the most critical challenges in IoT platforms and, hence, a lot of standards, drafts and research work has been proposed. There exist some security features within IoT protocols, however, that is not enough to fully secure the IoT systems. Several IETF standards have been proposed to offer security for such platforms including ACE, TLS/DTLS and many others discussed in this section. In addition, we presented some of the recent drafts discussing the challenges and threats to IoT security. It should be noted that IETF has a specialized group, called DTLS in constrained environments (DICE), for IoT security and is applying DTLS to all IoT applications. Furthermore, several recent research in providing IoT solutions have been discussed in this section.

## VIII. IoT Challenges

Despite the amount of work and standards on IoT, developing a successful IoT application is still not an easy task due to multiple challenges. These challenges include: mobility, reliability, scalability, management, availability, interoperability, cost and energy harvesting. In the following, we briefly describe each of these challenges.

### A. Mobility

IoT devices are supposed to move freely in the environment and, hence, change their IP addresses and connect to networks relative to their locations. Thus, routing protocols, such as RPL have to reconstruct the DODAG each time a node goes off the network or joins the network which adds a lot of overhead. In addition, mobility might result in a change of service provider which can add another layer of complexity due to service interruption and changing gateway.

### B. Reliability

For emergency response applications, it is very critical to keep the system perfectly working and delivering all of its specifications correctly. Hence, in IoT applications, the system should be highly reliable and fast in collecting data, communicating them and making decisions. Wrong decisions can lead to disastrous scenarios.

### C. Scalability

As millions and trillions of devices get connected in a single IoT application, scalability becomes a challenge that needs to be solved. Managing device distribution and functionalities is not an easy task. In addition, IoT applications should be tolerant of new services and devices constantly joining the network and, therefore, must be designed to enable extensible services and operations.

### D. Management

Even though several protocols to manage devices remotely were discussed, these protocols can not be applied to all IoT applications, and hence, management is still a big challenge. Providers need to manage faults, configuration, accounting, performance and security (FCAPS) of their interconnected devices.

### E. Availability

Availability of IoT platforms should guarantee both software and hardware availability for system users and service subscribers. Software availability means that the services are provided to the users, even when failures happen. Hardware availability means that the existing devices are easy to access and are compatible with various protocols. In addition, these protocols should be compact enough to be embedded within the constrained IoT devices.

### F. Interoperability

Interoperability means that heterogeneous devices and protocols need to be able to inter-work with each other. This is challenging due to the large number of different platforms used in IoT systems. Interoperability should be handled by both application developers and device manufacturers to deliver the services regardless of the platform or hardware specification used by the customer.

### G. Cost and complexity

Despite the relatively cheap prices of IoT devices such as sensors and smart transducers, it still costs too much to build an IoT application. Such complex integration of different protocols and standards makes IoT applications not available for general public usage. Reducing the cost and complexity is a massive challenge that needs to be solved.

### H. Power Harvesting

Power harvesting is still a challenge in IoT devices due to a lack of harvesting technologies for such small, resource constrained devices. Power is a critical issue in IoT as these devices need to last for years without battery changing and might be embedded in a body or environment which makes it difficult to change. Hence, collecting energy from motion or any other energy source and transforming it into stored energy seems to be a critical solution for such devices. However, such transformers and collection devices are still too weak to be applied to small devices due to their space and power needs.

I. Summary

This section discussed many current IoT challenges, including: mobility, reliability, scalability, and many others. Despite the amount of work in mobility, scalability, and management, enterprises still suffer these challenges. In addition to these challenges, security, as discussed before, is still a research challenge to be solved.

IX. Conclusion

In this paper, we have provided a comprehensive survey of protocols for IoT. A lot of those protocols have been developed and standardized by IETF, IEEE, ITU, and other organizations while many more are still in development. The discussion was brief due to the large number. Therefore, references for further information have been provided. The aim of this paper is to give an insight to developers and service providers about alternatives for different layers of protocols in IoT and how to choose among them. Through the paper, we classified our sections based on networking layers to: data link, network routing, network encapsulation, and session layers. At each layer, we presented most of the finalized standards and highlighted several drafts. In addition, we briefly reviewed IoT management protocols and discussed some of the existing security standards and work provided at different levels of standardizations. Finally, we discussed several challenges that still exist in IoT systems and that are being solved by researchers.